\documentclass[a4paper,11pt]{article}
\usepackage{jheppub} 
\pdfoutput=1
\usepackage{graphicx,array}
\usepackage[dvipsnames]{xcolor}
\usepackage{amstext}
\usepackage{amssymb}
\usepackage{slashed}
\usepackage{cancel}
\usepackage{subcaption}
\usepackage{booktabs}

\usepackage{pifont}

\definecolor{orange}{rgb}{1,0.5,0}
\definecolor{brown}{rgb}{0.59, 0.29, 0.0}
\definecolor{note_fontcolor}{rgb}{0.80078125, 0.80078125, 0.80078125}

\newcommand{\IR}{\text{ir}}
\newcommand{\UV}{\text{uv}}
\newcommand{\ir}{\text{ir}}
\newcommand{\uv}{\text{uv}}

\usepackage{tikz}
\usetikzlibrary{arrows,shapes.misc,decorations.markings,decorations.pathmorphing,positioning,intersections}

\usepackage{feynmp}
\usepackage{graphicx}
\usepackage{slashed}
\usepackage{tikz-feynman,contour}
\usepackage{tikz}
\usetikzlibrary{arrows,shapes.misc,decorations.markings,decorations.pathmorphing,positioning,intersections,patterns}

\tikzset{scalar/.style={dashed, > = latex, thick,
        decoration={markings,
            mark= at position 0.5 with {\arrow{#1}} ,
        },
        postaction={decorate}
    	},
    	gaugeboson/.style={decorate, decoration={snake},thick, segment length=3.2mm},
    	cross/.style={cross out, draw, 
         minimum size=2*(#1-\pgflinewidth), 
         inner sep=0pt, outer sep=0pt},
         gluon/.style={decorate, draw=black,
    decoration={coil,amplitude=4pt, segment length=5pt}}
}

\tikzfeynmanset{
  eftvert/.style={
    /tikzfeynman/dot,
    /tikz/minimum size=7pt,
  },
	every unparticle/.style={> = latex, double distance=2pt, decoration={markings,mark= at position 0.5 with {\arrow{#1}},},postaction={decorate}},
}

\tikzset{
	scalar/.style={dashed, > = latex, thick,decoration={markings,mark= at position 0.5 with {\arrow{#1}},},postaction={decorate}},
	gaugeboson/.style={decorate, decoration={snake},thick, segment length=3.2mm},
    fermion/.style={> = latex, thick,decoration={markings,mark= at position 0.5 with {\arrow{#1}},},postaction={decorate}},
    unparticle/.style={> = latex, double distance=2pt, thick,decoration={markings,mark= at position 0.5 with {\arrow{#1}},},postaction={decorate}}
}

\usepackage{subcaption}

\usepackage[bottom]{footmisc} 

\title{Confinement in de Sitter Space and the Swampland}
\author{Rashmish K.~Mishra}
\affiliation{Jefferson Physical Laboratory, Harvard University, Cambridge, MA 02138, USA}

\emailAdd{rashmishmishra@fas.harvard.edu} 

\abstract{
The relation between confinement scale $\Lambda_c$ of a gauge theory and the Hubble scale $H$ of the background quasi de Sitter space, required to satisfy the Festina Lente criteria ($\Lambda_c \gtrsim H$) from swampland principles, are investigated for a holographic theory using the Karch-Randall setup. A purely gravitational description admits unstable de Sitter extremum, and the requirement of Festina Lente translates into requirements of a perturbative gravitational dual, but prohibits an arbitrarily small hierarchy for other parameters fixed. With an additional scalar sector, the theory admits metastable de Sitter minima, and the requirement of Festina Lente imposes constraints on the parameters. For some range of parameters where Festina Lente could be constraining, the de Sitter minimum is very close to decay, thus invalidating the applicability. Further consequences of the finite lifetime of the dS extrema and the relation with related swampland conjectures are investigated, and no contradictions are found.
These results make case for the consistency and utility of the Karch-Randall setup, clarify the implications of Festina Lente for holographic theories, and connect Festina Lente to other swampland conjectures.
}

\begin{document} 
\maketitle
\flushbottom

\section{Introduction}
Seemingly consistent low energy effective field theories (EFTs) can be severely constrained by the requirement of a successful embedding into a quantum theory of gravity such as string theory. An interconnected web of conjectures has been put forth under the swampland program~\cite{Vafa:2005ui} (see~\cite{Palti:2019pca,vanBeest:2021lhn,Harlow:2022gzl} for a review), and has provided new perspectives on several theoretical~\cite{Lust:2022lfc} as well as phenomenological~\cite{Bedroya:2019snp,Bedroya:2020rac,Montero:2021otb,Montero:2022prj,Montero:2022jrc} research directions. Some of the most powerful conjectures, even if seen amply in string theory examples, can be equally derived by elegant arguments using black holes (BH), e.g. the weak gravity conjecture (WGC), absence of global symmetries and so on.

A qualitatively different condition, purely derived using BHs, but this time using a large Nariai BH in de Sitter (dS) space, is the so called Festina Lente (FL) condition~\cite{Montero:2019ekk}. Originally worked out for a $U(1)$ case, a straightforward implication of FL is that in a confining gauge theory that can be embedded in quantum gravity, the confinement scale must be larger than Hubble~\cite{Montero:2021otb}. This condition is fortuitously satisfied in the Standard Model (SM), and might have implications for beyond Standard Model (BSM) physics~\cite{Lee:2021cor,Ban:2022jgm}. 

The FL condition is special in three ways---one that it requires the mass to be \textit{more} than something, second that it can be stated \textit{without} any explicit mention of Newton's constant, and third that it must hold for \textit{every} particle in the theory, all of which are unlike WGC. 
It is natural to ask if taken on face value, what does this imply for EFTs, and in particular, are there EFTs where it can be studied, to gather evidence for this conjecture or otherwise, and constrain EFTs. In a direct approach, this would require studying confining gauge theories on a dS background, and is inherently complicated because it involves confinement. 

However, there is a well known EFT setup where questions related to gauge theories can be geometrized, usefully at least for strongly coupled gauge theories. In particular, the AdS/CFT duality, as applied to Randall-Sundrum (RS) framework with a UV and an IR brane~\cite{Randall:1999ee}, interprets the IR brane as the confinement scale~\cite{Rattazzi:2000hs}. The calculation of the location of the IR brane, usually done in the radion 4D EFT, calculates the confinement scale in the dual theory. This in itself would not be very useful for FL because in the original RS setup, the low energy 4D theory has zero cosmological constant, and hence FL is trivially satisfied, or does not apply. However, a closely related Karch-Randall (KR) framework~\cite{Karch:2000ct} relaxes the requirement of a zero cosmological constant in the 4D EFT. Using UV and IR branes whose tensions are different than those in the RS setup, one can have a 4D AdS or a dS theory. Holography therefore provides a way to study properties of confinement on a de Sitter background (e.g. see ref.~\cite{Marolf:2010tg} that focused on aspects of cosmological phase transition in dS). 

One way to think about the KR setup is from a consistent slicing of the 5D AdS space, given some energy in the UV. Depending on the tension of the UV brane being above or below a critical value, the consistent slicing is Anti-de Sitter (AdS) or dS. From a holographic point of view, the AdS vs dS cases are qualitatively different. The AdS slicing does not cut out all of the asymptotic boundary of the bulk, while dS slicing, like flat slicing, does. This means that for dS slicing, the interpretation of IR brane as a confinement scale in the dual theory can be justified---the energy cost of a separated quark anti-quark pair on the UV brane, dual to the area of the extended surface in the bulk, can not be indefinitely red-shifted once it approaches the IR brane, and can not escape to the asymptotic boundary in any direction (see~\cite{Maldacena:2003nj} for a review).

We therefore have a setup which naturally gives a non-zero Hubble and a non-zero confinement scale. Further, both of these can be easily calculated in the low energy 4D radion EFT. To study FL in this setup, with and without a stabilizing scalar sector, is the goal of this paper.

The organization of the rest is as follows. In Sec.~\ref{sec:review} we review the arguments of FL and the relevant ingredients of the KR setup. In Sec.~\ref{sec:potentials} we discuss the resulting potentials and the condition imposed by FL on the parameters, both with and without a stabilizing scalar sector. We conclude with some general remarks in Sec.~\ref{sec:conclusion}, focusing on the general structure of the potentials, the finite lifetime aspects of the extrema and the robustness of FL. The details of derivation of the effective potentials are given in App.~\ref{app:GW_Potentials}.

\section{The Ingredients}
\label{sec:review}
In this section we give a quick review of the arguments for Festina Lente (FL) conjecture, and some details of the Karch-Randall (KR) setup that are relevant for this work.
\subsection{Haste without Rush: Festina Lente conjecture}
By considering the phase diagram of Reissner-Nordstrom-de Sitter black holes in the charge-mass plane, the authors of ref.~\cite{Montero:2019ekk} argued that if a large Nariai black hole decays to dS, without resulting in a big crunch, this requires that every particle of mass $m$ and charge $q$ under a $U(1)$ gauge theory with gauge coupling $g$ must satisfy 
\begin{align}
    m^2 \gtrsim q\,g\,M_\text{Pl}\,H\:.
    \label{eq:FL-bound}
\end{align}
Unlike other swampland conjectures, this was argued without explicit string theory examples, however several arguments were given in support of the conjecture. The main point was that if there is even one particle that violates eq.~\eqref{eq:FL-bound}, a large charged Nariai BH discharges almost instantaneously, and since the charged Nariai BHs are more massive than the neutral ones, they collapse into a Big Crunch. This behavior would be against a thermodynamic expectation from taking the static patch of dS as a system with a temperature---dynamics in thermodynamical systems should describe thermal relaxation, and perturbations should equilibrate and disappear after some characteristic time. Further, the decayed BH would be in a region in the charge-mass plane which is unbounded and can be continuously deformed to arbitrarily large charge/mass solutions, which would have naked singularities. Additionally, these larger mass/charge solutions would have arbitrarily higher entropy, and this would be against a finite dimensional Hilbert space in de Sitter space.

In another work, ref.~\cite{Montero:2021otb} further studied the FL proposal, and pointed out the consequences for formal and phenomenological questions. It was pointed out that for a non-abelian gauge theory, there is a straightforward generalization of the FL proposal. Nariai BHs can be constructed for non-abelian gauge groups by embedding the standard Nariai solution in the Cartan of the non-abelian gauge group. Since a non-abelian gauge group automatically contains massless states, the gluons, the only way to satisfy FL conjecture would be that the gauge fields are either confined or Higgsed above the Hubble scale.
For a purely non-abelian gauge sector without any matter, FL implies
\begin{align}
    \Lambda_c \gtrsim H\;,
\end{align}
where $\Lambda_c$ is the confinement scale of the gauge theory, and $H$ is the Hubble constant. Taken on face value, this is a remarkable requirement, and seems to require a relation between two quantities that a priori have nothing to do with each other---$\Lambda_c$ is a purely QFT quantity, external to the Hubble of the background dS spacetime it is placed on. As an assurance, this is satisfied by several orders of magnitude in the Standard Model (SM). The precise $\mathcal{O}(1)$ numbers involved in the relation are not specified, and we will apply the conjecture with that in mind. There are two immediate comments in order. 

A well known bound on \textit{free} particles with mass $m$ and spin $s$ in de Sitter space is the Higuchi bound $m^2 \geq s(s-1)H^2$ in $d=4$ spacetime dimensions~\cite{Higuchi:1986py}. In a confining theory, one generically expects the low energy spectrum to contain states of mass of the order of confinement scale, and spin 0,1,2 or more (e.g. the stress-energy tensor acting on vacuum will generate a massive spin-2 state), so one could imagine that Higuchi bound must imply some kind of relation between $\Lambda_c$ and $H$. The Festina Lente condition is however more general than the Higuchi bound, since it also applies for spin $s=0,1$, and it does not say anything about the states being free particle states. It is to be noted that the underlying physics behind the two conditions is entirely different---Higuchi bound is derived by unitarity conditions on the free particle Hilbert space of a QFT in dS and has nothing to do with gravity, while FL is motivated by statements about black holes, that necessarily need gravity. 

A non-zero Hubble has a thermal interpretation, with a temperature proportional to the Hubble, since the accelerated observers experience Unruh radiation. If the confinement scale was below Hubble, this would imply that the phenomena of confinement would be shielded---all consistent QFTs would have the deconfined phase as the stable phase. The fact that FL requires this not to be true has an important implication from the point of view of the cosmological history of our universe. Our universe may be described by several isolated sectors that interact weakly with our sector, the SM, and this idea is motivated by the likely existence of dark matter and dark sectors that house them. Confining gauge theories are prototypical examples of such hidden sectors, and if so, the FL condition suggests we are likely to have many phase transitions in the history of our universe, as the temperature drops and crosses the confinement scale of a given sector. Assuming these sectors are somewhat populated after reheating, if some of the phase transitions are first order, the resulting gravitational waves might be detectable. 

\subsection{Bent without Dent: Karch-Randall setup}
5D AdS space with codimension one branes, studied under Randall-Sundrum brane worlds, have had a wide application in BSM physics, cosmology, KKLT construction, AdS/BCFT and recently in BH information paradox related questions (see ref.~\cite{Agrawal:2022rqd} for a brief overview). The original proposal involved a slice of 5D AdS Poincare Patch between a positive tension UV brane and a negative tension IR brane, whose tensions were equal in magnitude and to a critical value. The 4D induced geometry on the branes was flat, and this was suited to the phenomenological applications. 
Generalizing this requirement, solutions to Einstein equations with bent domain walls as branes were constructed~\cite{DeWolfe:1999cp,Kaloper:1999sm,Kraus:1999it,Garriga:1999bq} and the spectrum of the fluctuations around these solutions were studied in subsequent work~\cite{Karch:2000ct}.
Depending on whether the tension is more (less) then the critical value, the induced geometry on the brane was shown to be dS (AdS). For the dS case, one can use a coordinate system
\begin{align}
    ds_5^2 &= \sinh^2{r}\left(ds_{dS_4}^2 + dr^2\right)\;,
    \label{eq:5D-metric-dS-slice}
\end{align}
where a dS brane is at a fixed $r = r_*$, which is set by the tension on the brane~\cite{DeWolfe:1999cp}:
\begin{align}
    r_* = \coth^{-1}{\left(\left|\lambda\right|/\lambda_c\right)}\,
\end{align}
where $\lambda$ is the tension of the dS brane, and $\lambda_c$ is the critical value. The curvature and the Hubble parameter of a dS brane at $r=r_*$ is given as
\begin{align}
    H^{-1} = \ell = \sinh{r_*}\:.
\end{align}
The parameter $\ell$ is related to the effective cosmological constant on the brane
\begin{align}
    \Lambda = 3/\ell^2\:.
\end{align}
With one dS brane, the bulk geometry develops a horizon, the so called AdS Rindler horizon. The temperature of this horizon can be obtained by requiring smoothness of geometry near the horizon in the Euclidean picture with time direction compactified. In the coordinate system of eq.~\eqref{eq:5D-metric-dS-slice}, the horizon is at $r=0$.

For the dS case, one can have two dS branes---one with positive tension (more than the critical value), the UV brane, and one with negative tension (more in magnitude than the critical value, and less than the positive tension brane), the IR brane. In a recent work, a 4D effective action was written for the two dS brane scenario~\cite{Karch:2020iit}, which involved 4D Einstein gravity and a scalar $\varphi$, the radion. It is this scenario which is relevant for us, and which we will use in the next section. Note that the second brane can be put anywhere before the horizon, as long as its tension is chosen appropriately. The bulk spacetime terminates at the second brane, so the horizon is never reached.

By AdS/CFT, the KR setup with two dS branes is dual to a holographic CFT with a UV cutoff (corresponding to the UV brane), that confines at some lower scale (corresponding to the IR brane). As the CFT is put on a background with a non-zero Hubble, this affects the confinement, and there is additional contribution to the vacuum energy from confinement. All of this is captured in the 4D radion EFT. There is a non-trivial contribution to the radion action from the curvature, which requires redefining the radion properly. The scale of confinement $\Lambda_c$ and the effective Hubble $H$ are obtained from the potential of the redefined radion---as its extremum and the value at the extremum respectively. 

So far the discussion has only considered 5D Einstein gravity and its dimensional reduction to 4D. We can add a 5D scalar to this. In the dual picture this corresponds to adding scalar deformations to the CFT, which can change the low energy dynamics and confinement. However once again, this effect can be captured by the radion EFT once the effect of the scalar is included. In the next section we study the ratio $\Lambda_c/H$ for these cases, thereby checking what does FL say about holographic theories, and vice versa.
\section{Effective Potentials and Festina Lente}
\label{sec:potentials}
Equipped with the KR setup, we now consider the resulting 4D effective potentials for the radion. We first consider just the purely gravitational case, and later add a scalar sector as well. Investigating the relation between the extremum of the potential and the value of the potential at those extremum, we will be able to directly calculate the ratio $\Lambda_c/H$ and hence check the requirement of FL.
\subsection{Without a scalar sector}
We consider 5D AdS space with two branes with some given tension. The 4D effective action is given as (see~\cite{Chacko:2013dra,Karch:2020iit} for a derivation)
\begin{align}
    S = \frac{2 M_5^3}{k}\int d^4 x\sqrt{-g}
    \left(1-\varphi^2\right)\mathcal{R}
    -\frac{12 M_5^3}{k}\int d^4 x\sqrt{-g}
    \left(\partial\varphi\right)^2
    -\int d^4 x\sqrt{-g}\,
    V(\varphi)\:, 
\end{align}
where the potential is
\begin{align}
    V(\varphi) = \delta T_\text{UV} - \delta T_\text{IR}\varphi^4
    \equiv \frac{4 M_5^3}{k} \left( \frac{3}{\ell_\UV^2} - \frac{3}{\ell_\IR^2}\varphi^4\right)\:,
\end{align}
and we have taken the signs of the term for both branes to be supercritical (dS). The second line is defined to connect to the notation of Ref.~\cite{Karch:2020iit}. Note that due to the $\varphi^2\mathcal{R}$ term, we are not in the Einstein frame. For a non-zero background curvature, as is the case of interest, this term effectively makes the Newton's constant vary. We can go to the Einstein frame which is more suited. In terms of a new field $X$, related to $\varphi$ as
\begin{align}
    \varphi = \tanh{X}, 
\end{align}
the action looks like
\begin{align}
    S = \frac{2 M_5^3}{k}\int d^4 x\sqrt{-g}
    \mathcal{R}
    -\frac{12 M_5^3}{k}\int d^4 x\sqrt{-g}
    \left(\partial X \right)^2
    -\int d^4 x\sqrt{-g}\,
    V(X)\:.
\end{align}
Note that for a canonical kinetic term, we need to define $\widetilde{X} = \sqrt{12M_5^3/k}\,X$, but we will work with $X$ to keep the expressions simpler.
In the Einstein frame, the potential $V(X)$ is given as
\begin{align}
    V(X) &= \left(\cosh{X}\right)^4\, V(\varphi)|_{\varphi = \tanh{X}}\:
    \nonumber \\
    &= \frac{12 M_5^3}{k}\left(\frac{1}{\ell_\UV^2}\cosh^4{X}-\frac{1}{\ell_\IR^2}\sinh^4{X}\right)\:.
    \label{eq:Potential_noGW}
\end{align}
\begin{figure}[h]
\centering
\includegraphics[width=10cm]{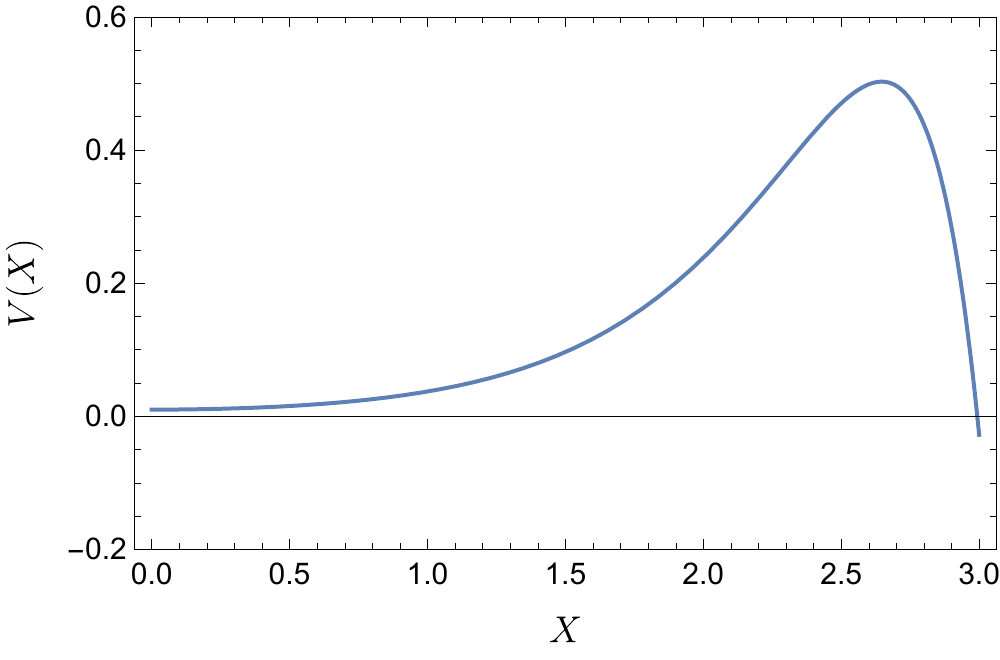}
\caption{\small{
Potential for $X$ for $\ell_\UV = 10, \ell_\IR = 9.9$, with $12 M_5^3/k = 1$. The potential has an unstable dS extremum at $X \sim 2.5$.
}}
\label{fig:Potential_noGW}
\end{figure}
The potential admits a dS extremum, albeit an unstable one (see Fig.~\ref{fig:Potential_noGW}). The potential is extremized at
\begin{align}
    \tanh{X_*} = \ell_\IR/\ell_\UV\:,
    \label{eq:location-of-extremum}
\end{align}
which is a maximum (as first pointed out in ref.~\cite{Chacko:2001em}). The value of the potential at the maximum is 
\begin{align}
    V_* = \frac{12 M_5^3}{k\ell_\UV^2}\frac{1}{\left(1-\ell_\IR^2/\ell_\UV^2\right)}\:.
    \label{eq:Potential-at-extremum}
\end{align}

Before proceeding to any discussion of FL, we can already ask if this resulting potential is consistent with swampland de-Sitter conjectures~\cite{Ooguri:2018wrx,Obied:2018sgi}, the relevant one being the refined de Sitter conjecture.
Using $M_\text{Pl}^2=2M_5^3/k$, and the canonically normalized field $\widetilde{X}$, we get
\begin{align}
    M_\text{Pl}^2\frac{d^2\,V}{d\widetilde{X}^2} &= -\frac{16 M_5^3}{k \ell_\uv^2}\frac{1}{\left(1-\ell_\ir^2/\ell_\uv^2\right)}\:.
\end{align}
Comparing this to the potential at the extremum, eq.~\eqref{eq:Potential-at-extremum}, we see that the refined de Sitter conjecture is trivially satisfied. 

We are now ready to check the requirement of FL to the dual theory. Using the fact that the vacuum expectation value (VEV) of $X$ sets the confinement scale in the dual theory, we have
\begin{align}
    \Lambda_c = \widetilde{X}_* = M_5 \sqrt{12 M_5/k}\,\tanh^{-1}{\left(\ell_\IR/\ell_\UV\right)}\:. 
\end{align}
The Hubble scale in the theory, set by Friedmann equations, is
\begin{align}
    H^2 = \frac{V_*}{3M_\text{Pl}^2} = \frac{2}{\ell_\UV^2\,\left(1-\ell_\IR^2/\ell_\UV^2\right)}\:,
\end{align}
where we used $M_\text{Pl}^2 = 2M_5^3/k$. The enhancement in Hubble, as $\ell_\ir/\ell_\uv \to 1$ is because in this limit, the branes are on top of each other, the 4D Planck scale is reduced and hence other quantities w.r.t. it must get enhanced. We therefore get
\begin{align}
    \Lambda_c/H = 2^{1/2}\,3^{1/2}\,\left(M_5/k\right)^{3/2}\,(k\,\ell_\UV)\sqrt{1-\ell_\IR^2/\ell_\UV^2}\,\tanh^{-1}{\left(\ell_\IR/\ell_\UV\right)}\:.
    \label{eq:FL_noGW}
\end{align}
For a well defined 5D gravitational description we need large $M_5/k$ (dual to the large $N$ of the dual theory), while $k\,\ell_\UV \gtrsim 1$, both of which go towards ensuring $\Lambda_c/H \gtrsim 1$.
The function $\sqrt{1-\ell_\ir^2/\ell_\uv^2}\tanh^{-1}{(\ell_\ir/\ell_\uv)}$ is bounded from above at $\ell_\ir/\ell_\uv\sim0.8$ for $\ell_\ir/\ell_\uv\in(0,1)$, the maximum value being approximately $0.66$. If $\ell_\ir/\ell_\uv$ is such that $\sqrt{1-\ell_\ir^2/\ell_\uv^2}\tanh^{-1}{(\ell_\ir/\ell_\uv)}$ is an $\mathcal{O}(1)$ number, the requirement of FL is
\begin{align}
    \left(M_5/k\right)^{3/2}\,k\ell_\UV \gtrsim 1\:.
\end{align}
This is also the condition for the dual gravitational description to be useful, and is necessarily satisfied. Hence FL is automatically guaranteed. Let us now go to the two extreme limits. If there is a large hierarchy such that $\ell_\IR/\ell_\UV \ll 1$, the right hand side of eq.~\eqref{eq:FL_noGW} simplifies, and the FL requirement is
\begin{align}
    \Lambda_c/H \sim \left(M_5/k\right)^{3/2}\,(k\,\ell_\IR) \gtrsim 1 \:,
    \label{eq:FL_cond_small_hierarchy}
\end{align}
which translates to a condition on the hierarchy:
\begin{align}
    \frac{\ell_\IR}{\ell_\UV} \gtrsim 
    \frac{1}{k \ell_\UV}\,\left(\frac{k}{M_5}\right)^{3/2}\:. 
\end{align}
Naively, for other parameters fixed, this forbids a very large hierarchy. However, in the limit of large $N$ of the dual theory, or equivalently, $M_5/k \gg 1$, one can make $\ell_\ir/\ell_\uv$ arbitrarily small. In the other extreme limit of the two branes approaching each other, $\ell_\ir/\ell_\uv \to 1$, eq.~\eqref{eq:FL_noGW} again simplifies to
\begin{align}
    \Lambda_c/H \sim \left(M_5/k\right)^{3/2}(k\,\ell_\uv)(1-\ell_\ir/\ell_\uv)^{1/2}\log \frac{2}{1-\ell_\ir/\ell_\uv}\:,
\end{align}
and approaches $0$ as $\ell_\ir/\ell_\uv \to 1$. Again in the limit of large $M_5/k$, one can make $\ell_\ir/\ell_\uv$ arbitrarily close to 1. However, for a given hierarchy $\ell_\ir/\ell_\uv$, this limits how large $N$ can be. For a small hierarchy, $\Lambda_c/M_\text{Pl} \sim \ell_\ir/\ell_\uv \ll 1$, using $N^2 \sim (M_5/k)^3$ and $k\,\ell_\uv \sim \mathcal{O}(1)$, eq.~\eqref{eq:FL_cond_small_hierarchy} implies $N \lesssim \Lambda_c/M_\text{Pl}$. This is the same condition derived in ref.~\cite{Kaplan:2019soo} using causality constraints and black hole entropy arguments.

These two limiting cases where FL condition was satisfied only in the limit of large N, are the situations when the IR brane approaches the bulk horizon ($\ell_\ir/\ell_\uv \to 0$) or the UV brane ($\ell_\ir/\ell_\uv \to 1$). The 5D picture is therefore very simple---the IR brane can be placed in the region between the UV brane and the horizon, and can be pushed all the way to the two extreme limits in the limit of large $N$. 

The KR setup is a very general construction, with general values of UV and IR brane tensions. In certain limits, it reduces to the RS geometry~\cite{Randall:1999ee}, the time-independent KR geometry~\cite{Karch:2000ct}, and the recent proposed time-dependent geometries~\cite{Karch:2020iit}. The effective Lagrangian, taken as a function of various parameters, is expected to be analytic, even if the solutions can have different local features and topology. This suggests that the conditions obtained here have a more general scope. Of course it may happen that a realization in string theory may come with additional conditions on the parameters, and will limit applying the conditions from FL on all the solutions.
\subsection{With a scalar sector}
So far, the discussion has focused on a potential which has a maximum as a dS. A natural question to ask is, what about meta-stable dS vacua? First of all what is the minimal ingredient that is needed to modify the potential, and what does FL say for the resulting dS. In the spirit of the older work of 4D Minkowski branes in AdS$_5$ with tuned tension that are stabilized after introducing a Goldberger-Wise (GW) scalar, one can attempt something similar here. The technical steps are definitely very similar.
The dual picture is that we are switching on various scalar deformations in the dual theory, which changes the features of confinement. As we will see, imposing FL requirement will constrain the parameters of the scalar sector in some cases. 

Once we include a 5D scalar $\Phi$ into the theory, we also need to specify the bulk and brane localized potentials for it. The brane localized potentials fix the boundary conditions for $\Phi$, and different choices lead to qualitatively different forms of potential for $X$. While there are several possibilities to consider, here we will only consider a quadratic bulk potential for $\Phi$ and limit to a small mass term for $\Phi$ (in units of the curvature). In this limit, the back-reaction on the geometry is small, and this suffices to generate interesting meta-stable dS minima. The case of including the back-reactions is complicated by the need to solve coupled equations, and we will not undertake it here, although interesting approaches have been suggested (e.g. see ref.~\cite{Kumar:2018jxz}). For the boundary conditions, we will consider three choices: DD (Dirichlet on IR, Dirichlet on UV), ND (Neumann on IR, Dirichlet on UV), and NN (Neumann on IR, Neumann on UV). The details of the procedure and the resulting potential for $X$ for various choice of boundary conditions are given in App.~\ref{app:GW_Potentials}.

Consider first the canonical choice of a Dirichlet boundary condition on both branes. The resulting potential for $X$, along with the contribution from before looks like
\begin{align}
V_\text{DD}(X) &= \frac{4 M_5^3}{k}
\left(
\frac{3}{\ell_\UV^2}\cosh^4{X}
-\frac{3}{\ell_\IR^2}\sinh^4{X} 
+\sinh^4{X}\,
\frac{(v_\IR - v_\UV \tanh^\epsilon{X})^2}{1-\tanh^{4+2\epsilon}{X}}
\right)\:.
\label{eq:Potential_GW_DD}
\end{align}
Here $v_\uv/v_\ir$ is the value of $\Phi$ on the UV/IR brane, $\epsilon = m^2/4k^2$ is related to the bulk mass $m$ of $\Phi$ and we are working to leading order in $\epsilon$. There are two things that can be noticed immediately. First, the contribution from the $\Phi$ field, the last term in the bracket, is manifestly positive. Further, it grows as the branes approach each other: $\tanh{X}\to1$. This is expected since the $\Phi$ field has to equal a different value on the two branes, so the two branes can't be on top of each other, as reflected by the growth of the third term. 

\begin{figure}[h]
\centering
\includegraphics[width=10cm]{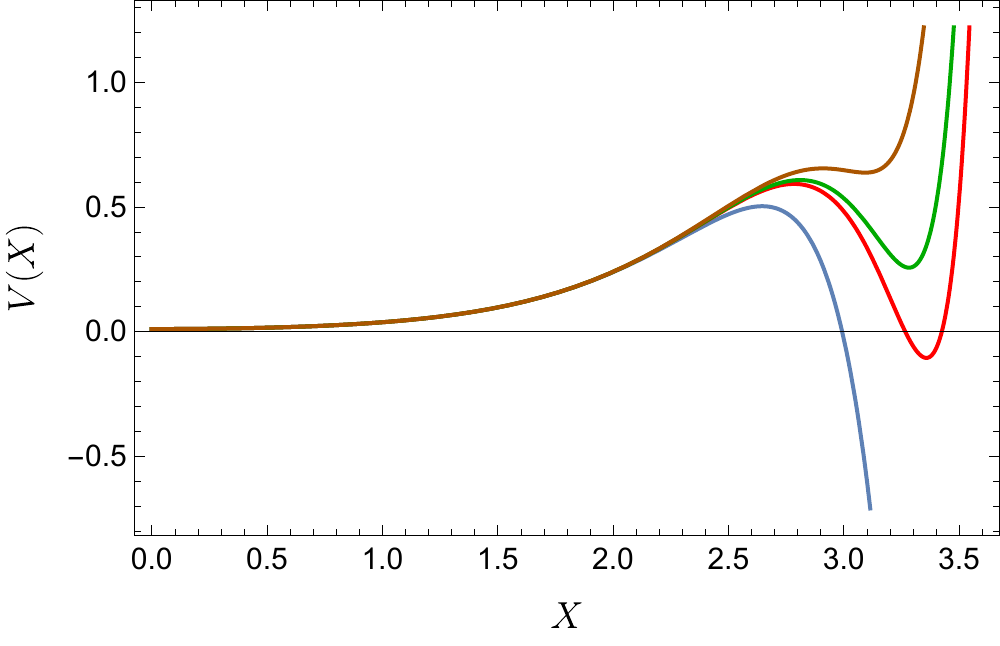}
\caption{\small{
Potential for $X$, in the presence of a scalar, with DD boundary conditions. The parameters chosen are $\ell_\UV = 10$, $\ell_\IR = 9.9$, $\epsilon = 10^{-2}$, $v_\IR = 10^{-3}$ and $1000\,v_\UV = 2$ (red), $2.05$ (green) and $2.15$ (brown), with $12 M_5^3/k = 1$. Also shown is the potential without the scalar contribution (blue), with the unstable extremum. The effect of the scalar dynamics is to increase $X_*$, and generate a meta-stable dS minimum.
}}
\label{fig:Potential_GW_DD}
\end{figure}

Since the potential now goes to positive infinity as opposed to negative infinity for large $X$, there is a possibility of a meta-stable dS minimum. Further, since the third term is positive, it effectively reduces $\ell_\IR$ so that the extremum shifts to larger values of $X$. The location of the minimum can be estimated, in the limit of $X\gg1, \epsilon\ll1$. Keeping the largest two terms in the large $X$ expansion, the minimum can be estimated to be as
\begin{align}
    e^{-2X_*} &\sim \frac{3}{16} \frac{\ell_\ir^2\,(v_\uv - v_\ir)^2}{1-\ell_\ir^2/\ell_\uv^2} \: .
\end{align}
The features of the potential are confirmed numerically for some choice of parameters in Fig.~\ref{fig:Potential_GW_DD}.

The effect of the scalar sector is to move $X_*$ to larger values. Without some fine-tuning of the parameters, one can entirely lose the dS minimum to an AdS minimum, as happens for one choice of parameters in Fig.~\ref{fig:Potential_GW_DD} (red curve). Further, only some choice of $v_\IR$ and $v_\UV$ do not require additional constraints from FL as compared to before---if $X_*$ increases and $V_*$ decreases as is the case for the green curve in Fig.~\ref{fig:Potential_GW_DD}, FL is not additionally constraining. This is however not a given in general, as shown by the brown curve in Fig.~\ref{fig:Potential_GW_DD}, though very quickly one runs into loosing the minimum altogether to a runaway. This suggests a relation between lifetime of dS and FL, and we will have more to say about it later.

Next consider the mixed boundary condition case---Dirichlet boundary condition on the UV brane, but a Neumann boundary condition on the IR brane. For this case we get the following form of the potential:
\begin{align}
V_\text{ND}(X) &= \frac{12 M_5^3}{k}
\left(
\frac{1}{\ell_\UV^2}\cosh^4{X}
-\frac{1}{\ell_\IR^2}\sinh^4{X} 
\right.
\nonumber \\
& \qquad\qquad\qquad\quad
\left.
+\frac83\alpha_\IR^2\,\sinh^4{X}\,
\left(
\frac{v_\UV\,\tanh^\epsilon{X}}{\alpha_\IR} -\frac{\left(1-\tanh^{4+2\epsilon}{X}\right)^2}{8}
\right)
\right)\:,
\label{eq:Potential_GW_ND}
\end{align}
where $v_\uv$ is the value of $\Phi$ on the UV brane, $\alpha_\ir$ is its derivative on the IR brane, $\epsilon$ is as before, and we again work in the small $\epsilon$ limit.

\begin{figure}[h]
\centering
\includegraphics[width=10cm]{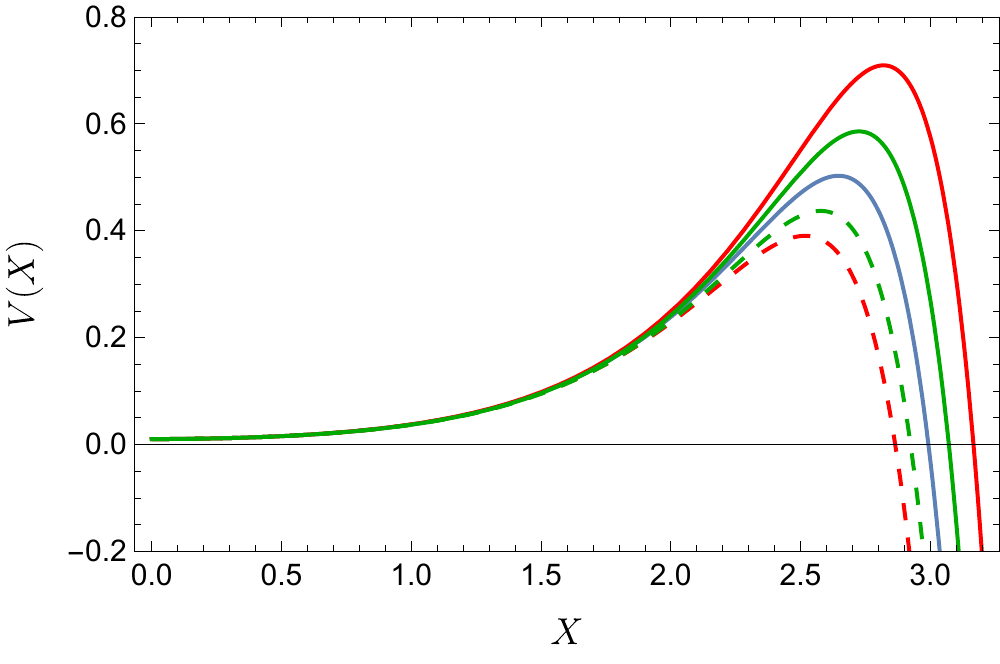}
\:\:
\caption{\small{
Potential for $X$, in the presence of a scalar, with ND boundary conditions. Left: the parameters are $\ell_\UV = 10$, $\ell_\IR = 9.9$, $\epsilon = 10^{-2}$, $(v_\UV, \alpha_\IR)$ = (0.03, 0.001) (red), (-0.03, 0.001) (red dashed), (0.0003, 0.05) (green), (0.0003, -0.5) (green dashed), with $12 M_5^3/k = 1$. Also shown is the potential without the scalar contribution (blue), with the unstable extremum. The effect of the scalar dynamics is to change $X_*$, but the unstable dS extremum stays unstable.
}}
\label{fig:Potential_GW_ND}
\end{figure}

Here, unlike the DD case, the extra contribution from the GW sector is not a perfect square, and therefore can be positive or negative depending on parameters. Also missing is the singularity as $X\to \infty$, since there is no conflict in the boundary conditions if the branes were to approach each other. One can therefore move the location of the unstable dS extremum to the left or right, or lose it altogether. The location of the maximum, in the large $X$ limit is
\begin{align}
    e^{-2X_*} &\sim \,\frac12\left(\frac{2v_\uv\alpha_\ir + 1/\ell_\uv^2 - 1/\ell_\ir^2}{(2+\epsilon)\,v_\uv\alpha_\ir - 1/\ell_\uv^2 - 1/\ell_\ir^2}\right)\:.
\end{align}
The features of the potential are confirmed numerically for some choice of parameters in Fig.~\ref{fig:Potential_GW_ND}.

\begin{figure}[h]
\centering
\includegraphics[width=10cm]{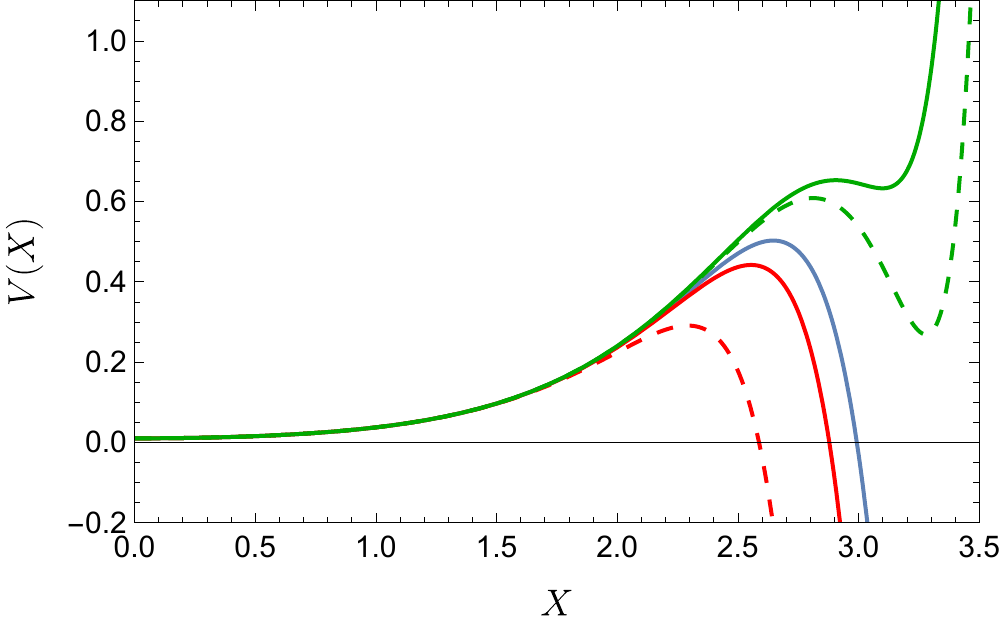}
\caption{\small{
Potential for $X$, in the presence of a scalar, with NN boundary conditions. The parameters chosen are $\ell_\UV = 10$, $\ell_\IR = 9.9$, ($\epsilon, \alpha_\UV, \alpha_\IR)$ = (0.03, 0.0001, 0.0001) (red), (0.03, 0.0001, 0.0005) (red dashed), (-0.031, 0.0001, 0.0001) (green), (-0.037, 0.0001, 0.0001) (green dashed), with $12 M_5^3/k = 1$. Also shown is the potential without the scalar contribution (blue), with the unstable extremum. The scalar dynamics in this case can give meta-stable dS as well as move the original dS maximum.
}}
\label{fig:Potential_GW_NN}
\end{figure}

We finally consider the NN case, the potential for which is given as
\begin{align}
    V_\text{NN}(X) &=
    \frac{12 M_5^3}{k}
    \left(
    \frac{1}{\ell_\UV^2}\cosh^4{X}
    -\frac{1}{\ell_\IR^2}\sinh^4{X} 
    -\frac{1}{3\epsilon}
    \frac{\left(\alpha_\UV^2+\alpha_\IR\,\tanh^{4+\epsilon}{X}\right)^2}{1-\tanh^{4+2\epsilon}{X}}
    \right)\:,
    \label{eq:Potential_GW_NN}
\end{align}
where $\alpha_\uv/\alpha_\ir$ is the derivative of $\Phi$ at UV/IR brane. This time, the scalar contribution is again a perfect square like the DD case, but its overall sign can be chosen by $\epsilon$. Further, it has the $1/\epsilon$ singularity, as well as the $\tanh{X}\to1$ singularity from mismatch of boundary conditions, which can make this term dominate for large $X$. We therefore expect to get metastable dS minimum for $\epsilon < 0$ (so that the potential becomes positive at large $X$) and simply move the unstable maximum to lower values for $\epsilon > 0$. The location of the metastable dS minimum, in the limit of $X\gg1, \epsilon\ll1$ and $\alpha_\uv = \alpha_\ir = \alpha$ can be estimated as
\begin{align}
    e^{-2X_*} & \sim \frac{3\,\alpha^2\,\ell_\ir^2}{-4\epsilon(1-\ell_\ir^2/\ell_\uv^2)}\:.
\end{align}
It is clear that for a real solution, we need $\epsilon < 0$. The features of the potential are seen numerically for some choice of parameters in Fig.~\ref{fig:Potential_GW_NN}.

\begin{figure}[h]
\centering
\includegraphics[width=10cm]{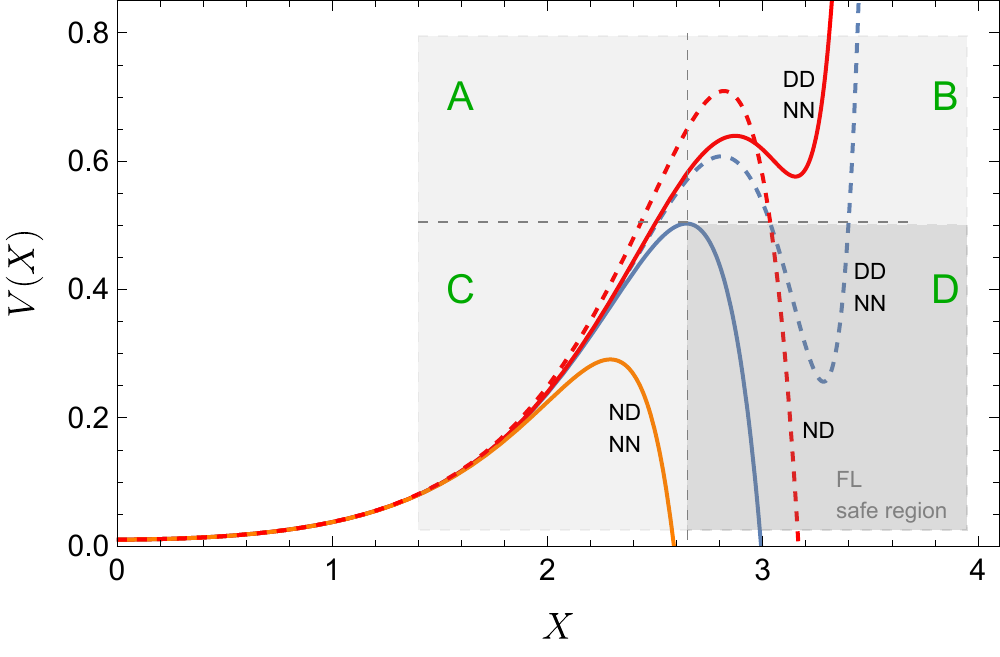}
\caption{\small{
The location of extrema of the radion potential for various choice of scalar sector parameters, as compared to the no scalar case (solid blue). The extrema can move in four possible directions (denoted by the four regions $A,B,C,D$), and the darker gray region $D$ is safe if the original extremum satisfied the FL condition. Requiring FL condition on the extrema in the other regions leads to constraints on the scalar sector. Also indicated are the boundary conditions that generically give rise to the typical potentials.
}}
\label{fig:Potential_GW_FLdirections}
\end{figure}

A pictorial summary of how the extrema move for various choices, as compared to the original extremum, is shown in Fig.~\ref{fig:Potential_GW_FLdirections}. For the 4 regions marked $A,B,C,D$, one can see that w.r.t. the original extremum (i.e. without a scalar sector), region $D$ is FL safe, region $A$ is FL unsafe, while regions $B,C$ can go either way. If an extremum moves to a region other than $D$, FL presents a constraint on the parameters of the EFT.

\subsection{Generic Feature of the Potentials and the Consequences for FL}
Having analyzed the potential for the radion without a scalar sector, and with the scalar sector for three choice of boundary conditions and a quadratic bulk potential, we can ask what is the general lesson towards FL. First of all, we saw that the no-scalar case and the ND case only admit a maximum, while the DD and NN case admit both a maximum and a minimum. Turns out one can understand this analytically.  Since these extrema appear at $X\gtrsim 1$, one can do a large $X$ expansion of the potentials. The generic form of the expansion is 
\begin{align}
    V(X) \: \underset{X\to \: \infty}{=} \: A\, e^{6X} + B\, e^{4X} + C\, e^{2X} + D\, \cdots \:.    
\end{align}
For the no-scalar and the ND case, $A=0$, so that the potential at large $X$ is simply a quadratic in $y=e^{2X}$ and at most can have one extremum. However, for DD and NN case, due to the $X\to\infty$ singularity, $A\neq 0$. The potential is now a cubic in $y=e^{2X}$ and admits a maximum and a minimum generically. 
For the cases in which one can obtain a dS minimum (DD and NN cases), we find $A>0$. Given that the potential should not blow up as $X\to 0$ (the limit of the IR brane falling to the asymptotic boundary, a sensible physical configuration), it is clear that a metastable dS minimum will require a sum of at least three exponentials, as seen explicitly.

One general lesson seems to be that whenever a meta-stable dS minimum appears, it is to the right of the original no-scalar maximum, and \textit{some} of them are at a lower height. This ensures that if the original maximum was consistent with FL, the meta-stable minimum will also be. However there are some minima that appear at a higher height, but they are very close to a runaway. It therefore seems that FL relates to the runaway behavior of the metastable dS minima. We will have more to say about this in the next subsection. The other consequence of a scalar sector has been to move the original unstable maximum, and this move is along a trajectory where FL can be constraining. For example, we saw that the maximum and its height either both increase or decrease, and that does not have to respect the original FL condition.

One feature of all the metastable minima is that they are at $\mathcal{O}(1)$ values of $X$. One can ask, what would it take to violate the FL requirement? Without a scalar sector, one can't violate FL since the requirement of a controlled gravitational description is tied to it. With the scalar sector, one can imagine getting a potential such that there is a meta-stable minimum at $X \ll 1$. For this, one can look at a small $X$ expansion of the potentials. For the potentials considered in this work, the generic form of this expansion is
\begin{align}
    V(X) \:\underset{X \to \: 0}{=} \:a + b\, X^2 + 
    \cdots \: .
\end{align}
In this expansion, one has to be careful about terms of the form $\tanh^\epsilon{X}$, which is not small, even if $X\ll1, \epsilon\ll1$. However, to the order we care, these terms do not show up. One can explicitly check that for no-scalar, DD and ND cases, $a = 1/\ell_\UV^2 > 0$ so that the potential starts at a positive value. This means there is a possibility of a dS minimum, if the potential can come down, which needs $b < 0$ (the potential should later go up due to next terms). However, for these cases, $b=2/\ell_\UV^2$, so there is an obstruction already. For the NN case, a similar arguments hold. The first two coefficients are $a^\text{NN} = (1/\ell_\UV^2 - \alpha_\UV^2/\epsilon)$ and $b^\text{NN} = 2(1/\ell_\UV^2 - \alpha_\UV^2/\epsilon) = 2\,a^\text{NN} $, and a dS minimum is again not possible.
 
This argument is only suggestive, since one can imagine several terms in the expansion conspiring to generate a minimum, but that requires $X^2 \sim X^4 \sim X^6$, and is generically not going to give an arbitrarily small $X$. Note that the constant term in the potential can not be changed arbitrarily by adding an overall constant, since the potential in the Einstein Frame does not have this freedom.

As we have argued using the small $X$ expansion, the potentials that arise seem very robust against having a minimum at very small $X$. We saw in the purely gravitational case that in the limit of large $N$ of the dual theory, we were able to get a maximum at very small $X$. If the potential with a scalar sector has to keep the extrema location but go from being unstable to stable, this needs a big back-reaction from the scalar sector on the geometry, as pointed out in ref.~\cite{Kumar:2018jxz}, which would also change the Hubble in the process. Since we are working in the limit of small back-reaction, we are not able to get such a minimum. It would be interesting to explore the FL condition in a setup where the back-reaction is included, and to see if the conclusion from the no scalar case still holds, that the now stable IR brane can be pushed all the way to the would be back-reacted horizon location. 

\subsection{Finite Lifetime of the Extrema}
A generic feature of all the extrema that result in this setup, whether maxima or minima, are that they are all finitely lived. Since this is obtained in a theory of gravity, this should not be surprising. This however immediately raises the question of whether the FL condition gets relaxed due to this. Further, it also suggests a possible connection, and at the very least, a comparison, with the Trans-Planckian Censorship Conjecture (TCC) conjecture~\cite{Bedroya:2019snp,Bedroya:2019tba}, which limits how long a dS extremum can live.

To understand when FL applies, we need to recall that the condition comes from prohibiting Nariai BHs to discharging to a singular spacetime. We therefore have three time scales to consider---$t_\text{dS}$ which is the life-time of the dS extremum, $t_\text{BH}$ which is the time in which a Nariai BH typically discharges, and $t_\text{TCC}$, which limits how long a dS space can exist. For FL to apply, we need
\begin{align}
    t_\text{dS} > t_\text{BH}\:,
\end{align}
so that the dS extremum still exists before the BH discharges. TCC on the other hand requires
\begin{align}
    t_\text{dS} < t_\text{TCC} = H^{-1}\,\log\left(\frac{M_\text{Pl}}{H}\right)\:.
\end{align}
The BH discharge time was estimated in~\cite{Montero:2019ekk,Montero:2021otb}, and is given as
\begin{align}
    t_\text{BH} \sim (qE)^{-1/2} = \left(q\,g\,H\,M_\text{Pl}\right)^{-1/2}
    = H^{-1}\left(\frac{H}{g\,M_\text{Pl}}\right)^{1/2} 
    = \left(\frac{1}{g^2\,V}\right)^{1/4}\:,
\end{align}
where in the last few terms, $q$ is set to 1 for non-abelian case. The gauge coupling $g$ appearing in the above is of the holographic dual theory, and is generically small.\footnote{For the specific example of $\mathcal{N}=4$ SYM, dual to $\text{AdS}_5\times S^5$, $g^2N = \lambda_\text{`t Hooft} = \left(M_\text{Pl}/M_\text{string}\right)^4 \gg 1$~\cite{Aharony:1999ti}. Requiring $N\to\infty$ effectively means $g\ll 1$ for finite (large) $\lambda_\text{`t Hooft}$.} However, using the magnetic WGC one can argue that the cutoff of the theory is $\Lambda_\text{cut-off} \sim g\,M_\text{Pl}$ and $H \lesssim \Lambda_\text{cut-off}$, so that, generically,
\begin{align}
    t_\text{BH} < t_\text{TCC}\:.
\end{align}
The magnetic version of FL, which limits how small $g$ can be~\cite{Montero:2021otb}, also gives the same result. This effectively means that for FL to apply, we need $t_\text{BH} < t_\text{dS} < t_\text{TCC}$.

Consider first the case of the unstable extremum, which arises in the purely gravitational and the ND case. The time scale for the field to fluctuate away from the top of the hill can be estimated by the Fokker-Planck equation that gives the time-dependent probability for field to be in a certain range. To leading order, the time scale is independent of the field value (e.g. see~\cite{Rudelius:2019cfh}), and is given as
\begin{align}
    t_\text{dS}^\text{top} = \frac{3H}{\left|V''\right|}\:.
\end{align}
A quick way to estimate this time is to displace the field from the top of the hill by order $H$ and calculate the time it takes to violate one of the slow roll conditions, e.g. $\eta$, as it rolls down. The condition for FL to apply becomes
\begin{align}
    \frac{M_\text{Pl}^2\,\left|V''\right|}{V} \lesssim 3^{-1/4}\,g^{1/2}\,\frac{M_\text{Pl}}{V^{1/4}}\:.
    \label{eq:FL-Cond}
\end{align}
Naively this would be a very strict requirement, and against the refined dS conjecture, for sufficiently small $g$ as required of holographic theories. However, the magnetic versions of WGC~\cite{Arkani-Hamed:2006emk} and FL~\cite{Montero:2021otb} both require that\footnote{Note that this implies a maximum value for $N$ in the dual theory, for a fixed `t Hooft coupling $\lambda$, which is not surprising given that dS has a finite dimensional Hilbert space interpretation.} 
\begin{align}
    g^2 \gtrsim \frac32\,\left(\frac{H}{M_\text{Pl}}\right)^2 \equiv g_\text{min}^2\;,
\end{align}
so that the condition for FL to apply, eq.~\eqref{eq:FL-Cond} becomes (using $3M_\text{Pl}^2\,H^2=V$)
\begin{align}
    \frac{M_\text{Pl}^2\,\left|V''\right|}{V} 
    \lesssim \left(\frac{g}{g_\text{min}}\right)^{1/2}\left(\frac{3}{2}\right)^{1/4}\:.
\end{align}
Note that this condition is consistent with the refined dS conjecture. For the explicit potential in the no scalar case, eq.~\eqref{eq:Potential_noGW}, $M_\text{Pl}^2\,V''/V = 4/3$. The condition for FL to apply is simply to have the gauge coupling slightly larger than its minimum value:
\begin{align}
    \frac{g}{g_\text{min}} \gtrsim \frac{16}{9}\left(\frac23\right)^{1/2} \approx 1.45\, \:.
\end{align}
For the maxima that appear with a scalar sector, it is straightforward to calculate this condition, and will be a constraint on the parameters of the scalar sector, for other parameters fixed.

Next consider the metastable dS minima that appear for the DD and NN cases. These extrema are classically stable, but quantum tunnelling effects give them a finite lifetime. Estimating the lifetime, given the explicit potential, is straightforward, though not tractable analytically. We can still make some non-trivial observations. First of all, if the dS minimum is in a ``FL safe region'' (region $D$ in Fig.~\ref{fig:Potential_GW_FLdirections}), i.e. the minimum is located at a larger value of $X$ and has a lower height, it already satisfies FL if the original maximum does. The lifetime of this dS minimum however needs to be larger than the BH discharge time, as discussed earlier, for FL to apply. 

When the minimum is in a ``FL unsafe region'', i.e. located at a larger $X$ but also has a higher height (e.g. region $B$ in Fig.~\ref{fig:Potential_GW_FLdirections}), something interesting happens. As seen from Figs.~\ref{fig:Potential_GW_DD} and~\ref{fig:Potential_GW_NN}, the barrier separating the dS minimum from the decompactification direction ($X \to 0$) gets small, and eventually the minimum is lost. The tunnelling rate in general depends on the height and the width of the potential barrier. There are two decay modes for the dS vacuum---the Coleman-De Lucia (CDL) instanton~\cite{Coleman:1980aw} from quantum tunnelling, and the Hawking Moss (HM) instanton~\cite{Hawking:1981fz} from thermal fluctuations. Denoting the height of the potential at the minimum/maximum by $V_0/V_1$, we can argue for the parametric dependence of the probability of decay on $V_0, V_1$. Since the CDL instanton is related to quantum tunnelling under the barrier, the probability for dS to decay from a CDL instanton is strongly enhanced as $V_0 \to V_1$. 
 
For a sufficiently small barrier, the CDL decay mode does not apply, but instead the HM instanton takes over. This instanton can be understood as a thermal effect, from the non-zero temperature of dS. The field fluctuates to the top of the hill, and rolls classically afterwards. Once again, as $V_0 \to V_1$, this effect is strongly enhanced. The lifetime of dS minimum in the FL unsafe region is therefore strongly reduced. 
Even though these estimates are crude, it is clear that the dS vacuum decays much before the typical BH discharge time, invalidating an application of FL criteria. 
It is intriguing that in a region where one can violate FL, the lifetime of dS decreases to prohibit applying FL altogether.
\section{Summary and General Remarks}
\label{sec:conclusion}
In this work we have applied the Festina Lente condition to a holographic confining gauge theory, using the dual language. We have used the Karch-Randall setup with two codimension one supercritical branes in 5D AdS space. In the low energy 4D EFT where the degrees of freedom are the 4D graviton and a scalar radion, we derived the radion potential, with and without a scalar sector. The extrema of the potential are dual to the confinement scale of the dual gauge theory, and the value of the potential is related to the Hubble constant. This allowed calculating the ratio $\Lambda_c/H$ in a straightforward way. 

We found that in the setup without a scalar sector, the requirement of FL was automatically satisfied for the dual gravitational picture to be sensible. We also found that for other parameters fixed, this limited how low the confinement scale can be relative to the UV cutoff. 
With a scalar sector we studied several choice of boundary conditions, and we showed that in some cases, the potential admits a meta-stable dS minimum, while in other cases the original unstable maximum just moves. 
For many of the metastable dS minima, we found that the location of the minima are such that it is already consistent with the original FL requirement. When not, these minima were very close to a runaway, and have a very small lifetime, which relaxes the FL condition. Further, there are cases when the new extremum is located such that FL condition on the original extremum is not sufficient, and this in general constrains the parameters of the scalar sector. We argued that the obtained potentials do not allow a minimum at arbitrarily small $X$, and will need non-trivial back-reaction to accomplish this. We also showed that the minimum can move such that $H$ becomes large, but this decreased the lifetime of dS minimum, thus relaxing the FL condition, before eventually losing the dS minimum altogether. These results make the case for the KR setup as a consistent EFT where such swampland conjectures and their relation to other conjectures can be studied.

The findings in this work, even though done in a different setup, are consistent with the expectations in full 10D spacetimes, as pointed out in ref.~\cite{Montero:2021otb}. For example, by considering a D3 brane wrapper around $S^3$ at the tip of the throat in the Klebanov Strassler (KS) geometry, a ``Baryon'', the requirement of FL was shown to be satisfied when one is in the regime of validity of the supergravity description. This is very similar to what we found here for the purely gravitational setup. Further, it was also pointed out that by using several decoupled sectors, one can engineer violating FL. This seems to be similar in spirit to adding several terms in the scalar sector and creating a FL violating extremum. In another work, using different tools, and without swampland motivation, ref.~\cite{Buchel:2019pjb} studied chiral symmetry breaking in a cascading $\mathcal{N}=1$ $SU(N+M)\times SU(N)$ gauge theory in de Sitter space. By numerically solving for the dual background geometry in various phases of the theory, they concluded that the scale of chiral symmetry breaking $\Lambda$ is constrained, and is generically of the order of Hubble, with $\mathcal{O}(1)$ numbers.

The usage of holography to study FL opens direction for studying the effect of other features of a confining gauge theory on the FL condition. If the gauge theory has conserved currents, the dual gravitational theory has gauge fields in the bulk, which are expected to have a Casimir contribution to the vacuum energy and hence generate a contribution to the potential for $X$. Weakly gauging these conserved currents, which may be Higgsed or confined in the deep IR, can also be studied in a similar setup as here. The consistency of the FL bound with the Higuchi bound is intriguing, since on the surface they are based on entirely different physical principles---it would be useful to understand this connection better. Finally, it might be possible to use the dS extrema obtained here for inflationary model building, including the unstable extrema (e.g. see ref.~\cite{Bedroya:2019tba}). We hope to return to these directions in future.

\section*{Acknowledgements}
I would like to thank A. Bedroya, H. Geng, A. Karch, Q. Lu, M. Montero, G. Obied, M. Reece, R. Sundrum, and C. Vafa for discussions at various stages, and comments on the draft. This work was supported by the National Science Foundation under Grant No. NSF PHY-1748958 and NSF PHY-1915071.
%
\appendix
\section{Potentials from a Scalar Sector}
\label{app:GW_Potentials}
In this appendix, we collect the results for the potential generated for $X$ from a scalar sector. The scalar sector is simply a 5D field $\Phi$ with given bulk and brane potentials. These potentials fix a background value for $\Phi$ which depends on the location of the IR brane. Evaluating the scalar action on this background value gives the potential for $\varphi$, which is related to the potential for $X$ by Eq.~\eqref{eq:Potential_noGW}.
Consider the 5D metric parametrized by
\begin{align}
    ds_5^2 &= e^{-2r}ds_4^2+dr^2\,\qquad 0 \leq r \leq r_\ir\:,
\end{align}
where $r$ is the bulk direction, and $\varphi = \exp(-r_\ir)$ is the radion.\footnote{This parametrization, even though naively suited for flat branes, works equally well for non-flat branes. This was shown for the case of ``matched'' as well as ``mismatched'' branes in ref.~\cite{Karch:2020iit}. The consequence of working in this particular 5D coordinate system is that the effective action will not in the Einstein frame, and needs a Weyl rescaling in the end. Reassuringly, the location of maximum and the negative radion mass squared for the no scalar case, obtained using this method, match the results in literature.} Consider a field $\Phi$ with the action
\begin{align}
S = \int d^5 x \sqrt{G} \left(\frac12 G^{MN}\partial_M \Phi\partial_N \Phi - V_\text{bulk}(\Phi)\right)
- \int_\uv d^4x \sqrt{G_\uv} V_\uv(\Phi)
- \int_\ir d^4x \sqrt{G_\ir} V_\ir(\Phi)\:.
\end{align}
The equations of motion for $\Phi(r)$ are (denoting differentiation w.r.t. $r$ by a $'$)
\begin{align}
\Phi'' -4 \Phi' - \partial_\Phi V_\text{bulk}(\Phi) - \delta(r)\,\partial_\Phi V_\uv(\Phi) - \delta(r-r_\ir)\partial_\Phi V_\ir(\Phi) = 0 \: .
\end{align}
The delta function terms impose the following boundary condition (obtained by integrating in a small neighborhood of the delta function)
\begin{align}
2\Phi'(r=0) &= \partial_\Phi V_\uv(\Phi)\:,\qquad
2\Phi'(r=r_\ir) = - \partial_\Phi V_\ir(\Phi)\: .
\end{align}
The difference in sign between UV/IR comes from which direction the delta function is approached. We will choose $V_\text{bulk}(\Phi) = \frac12 m^2 \Phi^2$. Since the bulk action is quadratic, the evaluation of action on the solution vanishes except at the boundaries, and is given by $\varphi^4 \Phi' \Phi |_0^{\log \varphi}$. The brane localized potentials give additional contributions depending on their form. The bulk equation for $\Phi$ is homogeneous and is solved by
\begin{align}
    \Phi(r) = A \varphi^{\nu_+} + B\varphi^{\nu_{-}}\:,
\end{align}
where $\nu_\pm=2\pm\sqrt{2+m^2}$, and $A,B$ are fixed by boundary conditions.  
We consider the following form of the potential on the UV/IR branes:
\begin{align}
V_{\uv/\ir}(\Phi) &= 2 \alpha_{\uv/\ir} \Phi + \beta_{\uv/\ir} (\Phi - v_{\uv/\ir})^2
\end{align}
The advantage of this form is that one can obtain both Neumann and Dirichlet boundary conditions in certain limits, the former corresponding to $\beta \rightarrow 0$ and the latter to $\beta \rightarrow \infty, \alpha \rightarrow 0$. For finite $\alpha,\beta$ the resulting boundary conditions are ``mixed''.
It is straightforward (but tedious) to work out the potential for boundary conditions with general $\alpha, \beta$. 
To leading order in $\epsilon$, the potentials for the three limiting cases considered in this work are given as
\begin{align}
    V_\text{DD}(\varphi) &= 4 \varphi^4 \frac{\left(v_\ir - v_\uv \varphi^\epsilon\right)^2}{1-\varphi^{4+2\epsilon}}\:,
    \nonumber \\
    V_\text{ND}(\varphi) &= \varphi^4 \left(2\alpha_\ir\,v_\uv\,\varphi^\epsilon - \frac{\alpha_\ir^2}{4}(1-\varphi^{4+2\epsilon})\right)\:,
    \nonumber \\
    V_\text{NN}(\varphi) &= -\frac{\left(\alpha_\uv + \alpha_\ir \varphi^{4+\epsilon}\right)^2}{\epsilon(1-\varphi^{4+2\epsilon})}\:.
\end{align}
The potential for more general parameters can be easily obtained using similar steps.
\bibliographystyle{utphys}
\bibliography{references}
 
\end{document}